 \documentclass[prb,aps,twocolumn,superscriptaddress,showpacs,nofootinbib]{revtex4-2}
\usepackage{mhchem}
\usepackage{gensymb}
\usepackage{booktabs}
\usepackage{graphicx}
\usepackage{dcolumn}
\usepackage{bm}
\usepackage[dvipsnames]{xcolor}
\usepackage{float}
\usepackage{amsmath,amssymb}
\usepackage[colorlinks=true,linkcolor=black, citecolor=blue, urlcolor=blue, 
unicode=true,breaklinks=true]{hyperref}

\newcommand{\nro}{Na$_{3+x}$Ru$_{3-x}$O$_6$}

\usepackage{chemformula}[2014/04/08]
\setchemformula{kroeger-vink}

\begin{document}

\preprint{APS/123-QED}

\title{Defect control in the Heisenberg-Kitaev candidate material NaRuO$_2$}%

\author{Brenden~R.~Ortiz$^{\dagger}$}
 \email{ortiz.brendenr@gmail.com}
  \thanks{These authors contributed equally}
 \affiliation{Materials Department, University of California Santa Barbara, Santa Barbara, CA 93106, United States}%
 
\author{Paul~M.~Sarte}
 \email{pmsarte@gmail.com}
 \thanks{These authors contributed equally}
 \affiliation{Materials Department, University of California Santa Barbara, Santa Barbara, CA 93106, United States}%
 
 \author{Alon~H.~Avidor}
 \affiliation{Materials Department, University of California Santa Barbara, Santa Barbara, CA 93106, United States}%

\author{Stephen~D.~Wilson}
 \email{stephendwilson@ucsb.edu}
 \affiliation{Materials Department, University of California Santa Barbara, Santa Barbara, CA 93106, United States}%

\date{\today}

\begin{abstract}
 The combination of geometric frustration, extended hopping, spin-orbit coupling, and a disordered magnetic ground state make NaRuO$_{2}$ an attractive Heisenberg-Kitaev candidate material. Historically, NaRuO$_2$ has been a challenging material to produce, even in polycrystalline form. Here we present synthetic efforts that identify a propensity for Na$_\text{Ru}$ defects to form in NaRuO$_2$, revealing a full solid-solution between NaRuO$_{2}$ and disordered Na$_2$RuO$_3$. We report the synthesis of alloys along the \nro~solid solution and characterize changes in the bulk magnetization and electron transport as a function of Na-loading.  Our results highlight the importance of stoichiometry control in NaRuO$_{2}$ when investigating and interpreting this material's physical properties.
\end{abstract}

\maketitle

\section{Introduction}

Unambiguous experimental realization of a quantum spin liquid (QSL) state remains an enduring challenge \cite{Balents10:464,Zhou17:89,broholm20:367}. Characterized by a ground state featuring highly entangled spins exhibiting no long-range magnetic order, QSL states are born out of an intricate and often subtle interplay of comparable, often competing, energy scales and are thought to be quenched by relatively small perturbations. Thus, understanding and controlling crystalline disorder, structural distortions, chemical impurities, and intrinsic defects are critical challenges when developing QSL phenomenology in real materials.

\begin{figure}
\includegraphics[width=1\linewidth]{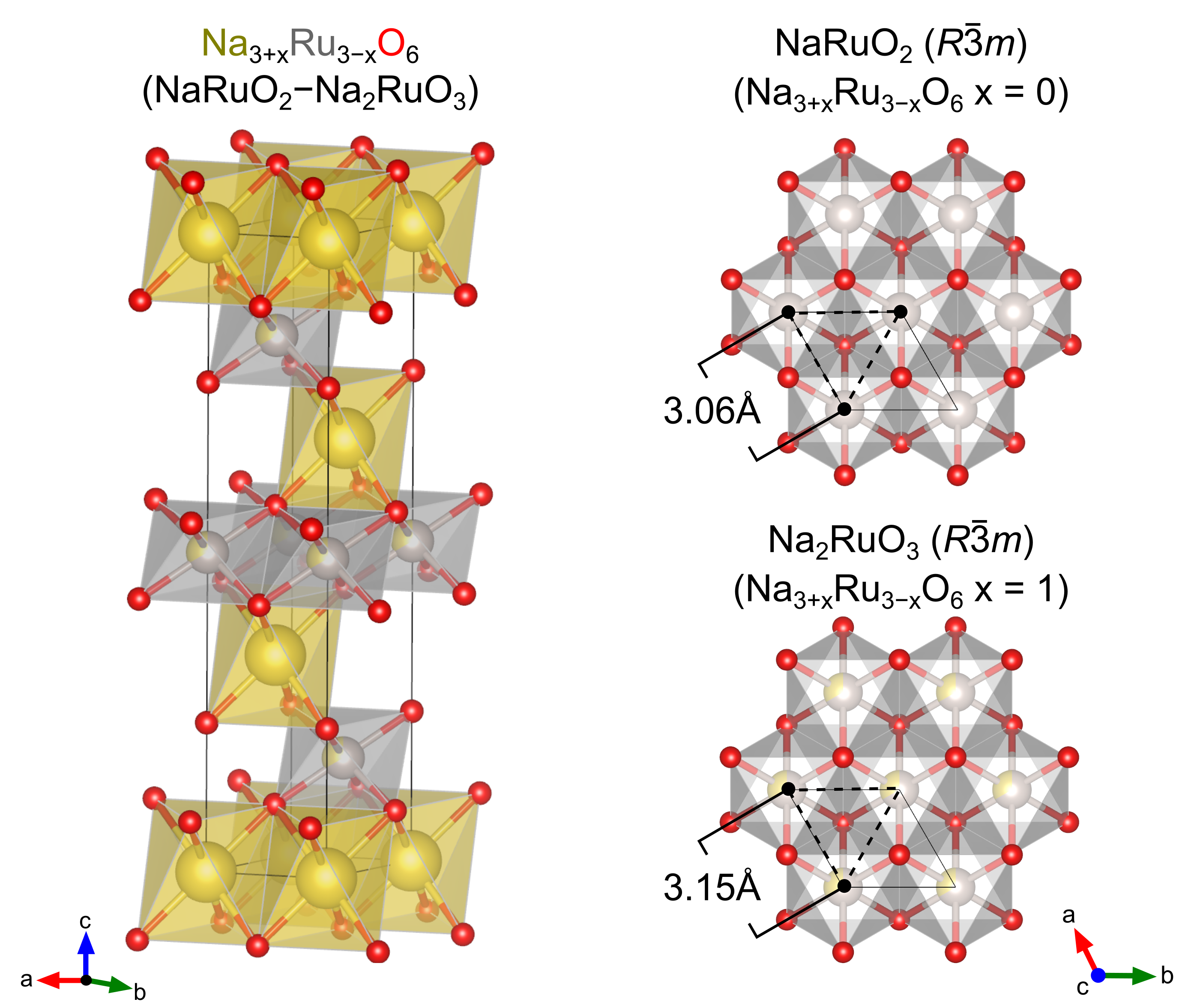}
\caption{Delafossite ($R\bar{3}m$) crystal structure assumed by the \nro~solid solution between the ternary end members NaRuO$_{2}$ ($x$=0) and disordered Na$_{2}$RuO$_{3}$ ($x$=1). \nro~forms a triangular sublattice comprised of edge-sharing Ru$^{3+}$ (4$d^{5}$) octahedra. Na-rich conditions overwhelmingly favor formation of Na$_\text{Ru}$ anti-site defects, diluting the Ru$^{3+}$ sublattice with nonmagnetic Na$^{+}$.}
\label{fig:Crystal}
\end{figure}

NaRuO$_2$ is a newly proposed, candidate QSL host that straddles a unique energy landscape -- one where Heisenberg-Kitaev interactions as well as extended exchange foster a native, quantum disordered ground state \cite{ortizNaRuO2}. NaRuO$_2$ is a member of the layered family of $AB$O$_2$ delafossite-like oxides, a larger family of $R\overline{3}m$ quasi-two-dimensional materials that support ideal antiferromagnetic triangular lattices on the $B$-site sublattice. Specifically, NaRuO$_2$ (Figure \ref{fig:Crystal}) features a triangular lattice of Ru$^{3+}$ ions separated by planes of Na$^+$. The edge-sharing RuO$_6$ octahedra place the Ru$^{3+}$ (4d$^5$) ions in a lightly trigonally distorted cubic crystal field. With appreciable spin-orbit coupling $\lambda$ and Coulomb repulsion $U$, the system is capable of supporting a half-filled $J_\text{eff}=1/2$ orbital. The result is a weak $J_\text{eff}=1/2$ Mott state with a disordered magnetic ground state and energetic antiferromagnetic interactions \cite{ortizNaRuO2}.

Despite lacking native chemical disorder such as that present in triangular lattice compounds like YbMgGaO$_{4}$ \cite{Paddison17:13,li19:2}, off-stoichometry and the resulting defects are a persistent concern among the alkali metal delafossite variants \cite{Dally17:459,clarke1998synthesis}. The typical culprit tends to be alkali-metal vacancies, whose presence is traditionally countered by the introduction of an excess of alkali precursors during growth. However, the historical precedent for alkali-vacancies as the dominant defect often neglects complex structure-defect-property relationships that can dominate in real systems -- NaRuO$_2$ is one such example.

In this work, we examine the defect chemistry of the Heisenberg-Kitaev candidate material NaRuO$_{2}$, mapping the Na--Ru--O phase diagram in the vicinity of NaRuO$_{2}$ to understand the extent and type of off-stoichiometry supported by the compound. We demonstrate the formation of a single solid-solution \nro~between the triangular lattice compound NaRuO$_{2}$ and the disordered honeycomb lattice compound Na$_{2}$RuO$_{3}$ \cite{mogare2004syntheses}, highlighting the tendency for NaRuO$_{2}$ to form Na-rich Na$_\text{Ru}$ defects. A combination of bulk magnetization and electron transport measurements highlight strong property changes as a function of Na-loading, highlighting the importance -- and more importantly -- the ability to control stoichiometry in NaRuO$_{2}$.

\section{Experimental Methods}

\subsection{Synthesis}

Polycrystalline members of the \nro~solid solution were synthesized using the same mechanochemical methods detailed in our recent work \cite{ortizNaRuO2}. Na$_{2}$O$_{2}$ beads (Sigma, 97\%), RuO$_{2}$ powder (Alfa, 99.95\%), and Na metal (Alfa 99.8\%) were 
combined in a pre-seasoned tungsten carbide ball mill vial and sealed under Ar. Due to the volitility of Na and and potential oxygen off-stoichiometry in RuO$_{2-x}$, adjustments are required to the nominal Na:Ru:O ratios. Specifically, both the compositions for Na$_2$RuO$_3$ and NaRuO$_2$ were empirically tuned to yield phase-pure compositions at Na$_{1.07}$(RuO$_2$)$_{1.13}$(Na$_2$O$_2$)$_{0.70}$ (Na$_{2.0}$Ru$_{0.9}$O$_{3.0}$) and Na$_{1.07}$(RuO$_2$)$_{1.37}$(Na$_2$O$_2$)$_{0.37}$  (Na$_{1.0}$Ru$_{0.8}$O$_{2.0}$) respectively. Using a combination of excess Na metal, Na$_2$O$_2$ and RuO$_2$, we iteratively narrowed down the single-phase region of the NaRuO$_2$--Na$_2$RuO$_3$ alloy, adjusting the compositional vectors until secondary phases were eliminated. All alloys were generated through a subsequent linear interpolation of the \textit{tuned} compositions of Na$_2$RuO$_3$ and NaRuO$_2$. Empirical tuning and interpolation is essential, as the compensating ratio of Na:Ru:O that yields phase pure NaRuO$_2$ is not the same as the compensation required for Na$_2$RuO$_3$. 

The resulting mixture was milled for 60~min in a Spex 8000D Mixer/Mill using four 7.9~mm tungsten carbide balls. The reaction generates a substantial amount of heat, and care must be taken with large sample volumes. The resulting precursor is confirmed amorphous by powder x-ray diffraction. The milled powder was then lightly ground in an agate mortar under Ar to disperse any agglomerates, sieved through a 100 micron sieve, and loaded into 2~mL alumina cylindrical crucibles (CoorsTek). In addition, a small portion of the milled powder was cold-pressed into 5~mm diameter pellets and buried within the powder bed. The crucibles were subsequently sealed under 1~atm of Ar in fused silica ampoules and placed within a 900$^{\circ}$C preheated furnace. Samples were annealed for 30~min and then immediately air-quenched before extracting powders under Ar. The final powders and sintered pellets are largely phase pure with trace amounts of Ru metal ($<$2~\%). Powders are black and moisture sensitive, with sensitivity increasing dramatically with additional Na content.

\subsection{Structural Characterization}
Phase purity was initially examined with powder x-ray diffraction (XRD) measurements at room temperature on a Panalytical Empyrean diffractometer (Cu K$_{\alpha_{1,2}}$) in Bragg-Brentano ($\theta$-$\theta$) geometry. \nro~powders were placed on a Si zero-diffraction plate under argon and capped with a 12~mm$\times$12~mm piece Kapton film to shield against atmospheric moisture.  Pawley and Rietveld refinements were performed using \texttt{TOPAS Academic} v6 \cite{Coelho}. Structural models and visualization utilized the \textsc{VESTA} software package \cite{Momma2011}. 

\subsection{Magnetization and Electron Transport Measurements}
Temperature dependent dc-magnetization data under zero-field-cooled (ZFC) and field-cooled (FC) conditions were collected on a 7~T Quantum Design Magnetic Property Measurement System (MPMS3) SQUID magnetometer. Samples were sealed in polypropylene holders under argon to minimize absorption of atmospheric moisture. Data was collected continuously in sweep mode with a ramp rate of 2~K/min in the presence of an external DC field of 1000~Oe. Isothermal dc-magnetization measurements at 2~K were collected continuously in sweep mode with a ramp rate of 100~Oe/sec.

Resistivity measurements were performed on sintered pellets of \nro~that were sectioned into rectangular bars with approximate dimensions of 1$\times$2$\times$0.5~mm. Electrical contacts were made in a standard four-point geometry with contacts being made with a combination of gold wire and silver paint. Thermal contact and electrical isolation was ensured using layers of GE varnish and cigarette paper. The temperature dependence of the electrical resistivity was measured with the Electrical Transport Option (ETO) in a 9~T Quantum Design Dynacool Physical Property Measurement System (PPMS) using a drive current of 10 $\mu$A and drive frequency of 100~Hz. Data was collected continuously in sweep mode with a ramp rate of 2~K/min.

\section{Results \& Discussion}
\subsection{Synthesis \& Structure}

Motivated by the combination of strong spin-orbit coupling, the expanded nature of the Ru $d$-orbitals, and remnant Coulomb interaction effects, ruthenates have continued to garner substantial attention. Owing to the many stable oxidation states of Ru, the Na--Ru--O phase diagram is remarkably complex. Within a relatively narrow set of chemical potentials there are at least 7 reported Na--Ru--O ternary compounds: NaRuO$_2$ \cite{shikano2004naruo2}, NaRu$_2$O$_4$ \cite{shikano2004synthesis}, Na$_2$RuO$_3$ \cite{mogare2004syntheses}, Na$_3$RuO$_4$ \cite{regan2005isolated}, Na$_2$RuO$_4$ \cite{mogare2004syntheses}, Na$_{27}$Ru$_{14}$O$_{48}$ \cite{allred2011na27ru14o48}, and Na$_{3-x}$Ru$_4$O$_9$ \cite{regan2006structure}. 

NaRuO$_2$ is of particular interest due to the triangular sublattice of Ru$^{3+}$ and the potential applications as a QSL candidate material \cite{ortizNaRuO2}. Remarkably, a survey of adjacent phases to NaRuO$_2$ reveals the ``disordered'' ($R\bar{3}m$) polymorph of Na$_2$RuO$_3$ is structurally identical to NaRuO$_2$, except for the random dilution of the Ru$^{3+}$ triangular sublattice with nonmagnetic Na$_\text{Ru}$ defects. It is important to note that while Na$_2$RuO$_3$ can also crystallize in a ordered $C2/c$ monoclinic structure, it is not clear which phase is the thermodynamic ground state.

Such a relationship and the resulting potential for off-stoichiometry in NaRuO$_{2}$ is supported by a comparison the available crystallographic data. The original synthesis procedure reported for NaRuO$_2$ involves a three step decomposition process where: 1) \ce{Na2RuO4} was synthesized from a stoichiometric mixture of \ce{Na2O2} and \ce{RuO2}, 2) stoichiometric amounts of \ce{Na2RuO4} and Ru metal were mixed, dried, and sealed inside gold tubing, and finally 3) the mixture was heated at 1173~K for 12~h and then 1273~K for 120~h \cite{shikano2004synthesis}. This processing route produces material with lattice parameters [$a,c$] : [3.02~\AA, 16.49~\AA]. We have developed a new, rapid, mechanochemical route for the synthesis of NaRuO$_{2}$ \cite{ortizNaRuO2}, which is the method utilized in the present study. This processing route renders NaRuO$_2$ with lattice parameters [3.06~\AA, 16.18~\AA].

The difference observed in the $c$-axis lattice parameters reported in this work \cite{ortizNaRuO2} and prior work by Shikano et al. \cite{shikano2004naruo2} is substantial and noteworthy. One potential origin of this discrepancy is the impact of Na off-stoichiometry, which would naturally impact the interlayer spacing. Looking to the analogous titanate structure (Na$_{1-x}$TiO$_{2}$), detailed structural studies have identified a contraction along \textit{c} and an expansion in \textit{a} as Na-vacancies were eliminated and the composition approached nominal NaTiO$_2$ \cite{clarke1998synthesis}. We suggest that the smaller \textit{c}-axis lattice parameter of NaRuO$_{2}$ synthesized via the mechanochemical route presented herein are closer to the ideal 1:1:2 stoichiometry. This is further supported by our previous neutron powder diffraction refinement \cite{ortizNaRuO2}, which indicates that the \textit{tuned} NaRuO$_2$ composition is stoichiometric within the resolution of our measurement. The discrepancy between the prior report and our results suggests that off-stoichiometry and defect control are important factors in NaRuO$_2$.

Drawing inspiration from the thermoelectric community and the concept of ``phase boundary mapping" \cite{pbmortiz2019carrier, pbmohno2017achieving,pbmohno2018phase,pbmcrawford2018experimental}, we sought to map the phase space surrounding NaRuO$_2$. Wide swaths of the space immediately surrounding NaRuO$_2$ are dominated by 2-phase equilibria, which is unexpected if NaRuO$_2$ is a prototypical line compound. This is instead consistent with the formation of a large single-phase region or an extended alloy. Furthermore, NaRuO$_2$ shows an unusual proclivity to incorporate excess Na into the structure. Considering the structural similarity of disordered Na$_2$RuO$_3$, we suspected that a solid solution between NaRuO$_2$ and Na$_2$RuO$_3$ could exist. In support of this conjecture, synthesizing Na$_{2}$RuO$_{3}$ using the same synthetic conditions as NaRuO$_{2}$ results in the formation of disordered $R\bar{3}m$ Na$_{2}$RuO$_{3}$. This disordered Na$_{2}$RuO$_{3}$ polymorph persists after extended annealing and appears to be the stable structure under our processing conditions.

\begin{figure}
\includegraphics[width=1\linewidth]{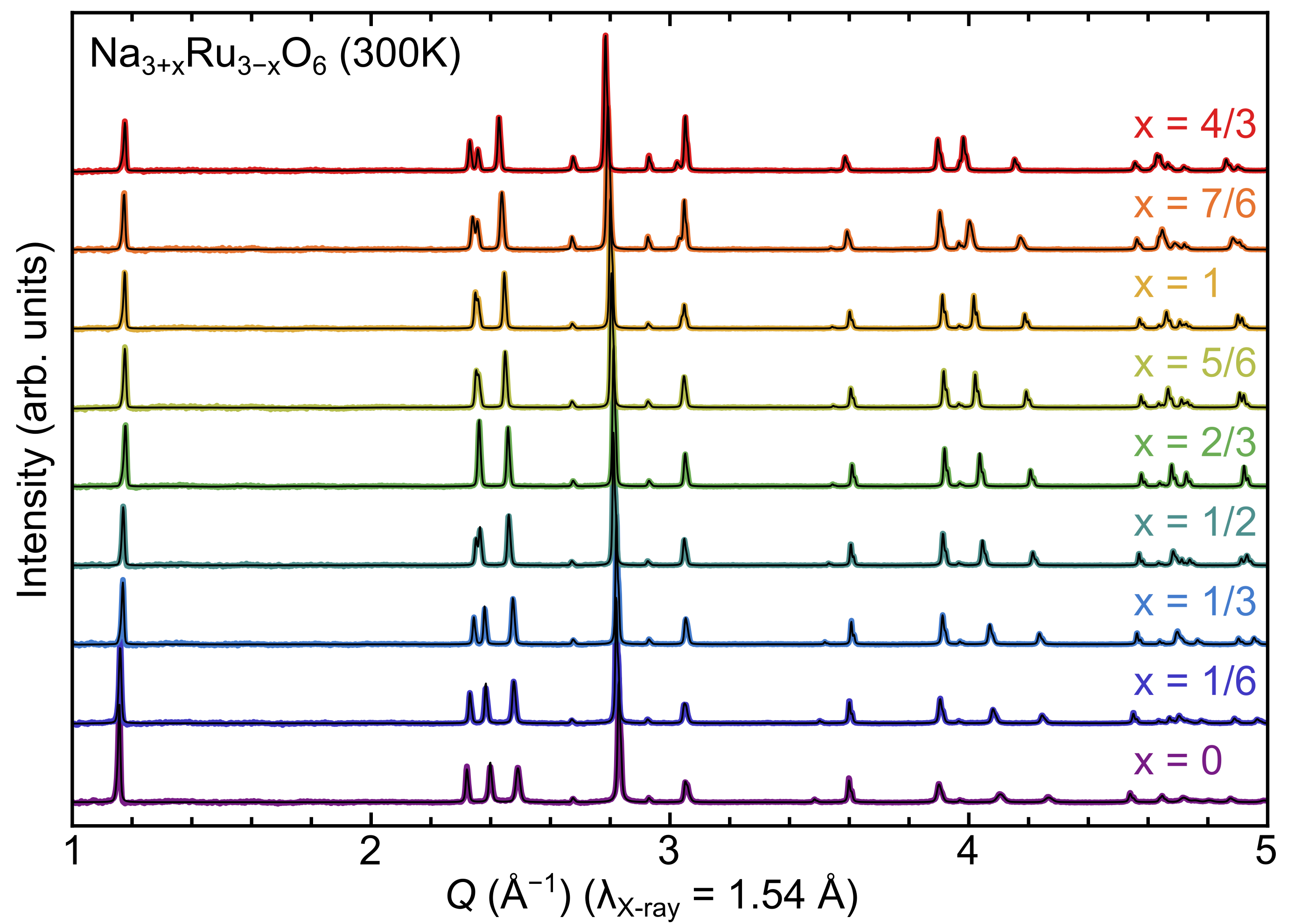}
\caption{X-ray patterns of Na$_{3+x}$Ru$_{3-x}$O$_6$ alloy series demonstrate Successful alloying of NaRuO$_{2}$ ($x$=0) and Na$_2$RuO$_3$ ($x$=1) through continuous shifts in the peak positions and intensities. Black traces indicate resulting Pawley refinements in the $R\overline{3}m$ structure. All samples up to $x$=1 are predominately phase-pure Na$_{3+x}$Ru$_{3-x}$O$_6$ with trace Ru metal. Samples extending beyond nominal Na$_{2}$RuO$_{3}$ ($x$=1) exhibit increased Ru formation, suggesting a geometrical shift in the single-phase boundary.}
\label{fig:Scattering}
\end{figure}

\begin{figure}
\includegraphics[width=1\linewidth]{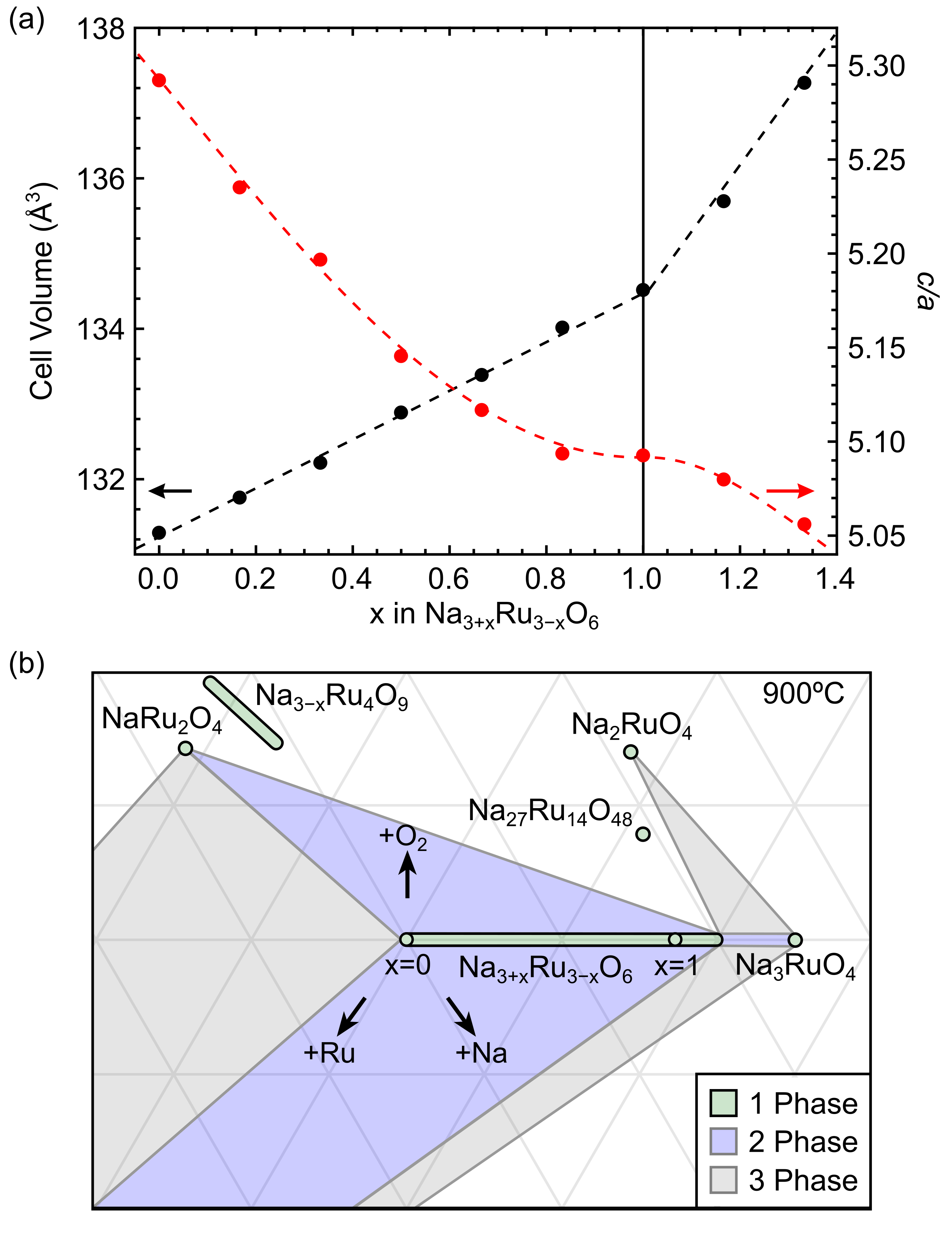}
\caption{(a) Compositional dependence of the unit cell volume (black) and the c/a ratio (red) for the \nro~solid solution extracted from Pawley refinements of room temperature pXRD data. (b) Tentative processing ternary phase diagram schematic at 900$^{\circ}$C isotherm for Na--Ru--O space surrounding the \nro~solid solution.}
\label{fig:Vegard}
\end{figure}

To verify the solid-solution hypothesis, a series of samples ranging from NaRuO$_{2}$--Na$_{2}$RuO$_{3}$ were synthesized. For the sake of convenience, we will refer to the series using the renormalized stoichiometry Na$_{3+x}$Ru$_{3-x}$O$_6$ where the end members of $x$=0 and $x$=1 correspond to nominal NaRuO$_{2}$ and Na$_{2}$RuO$_{3}$, respectively. As illustrated in Fig.~\ref{fig:Scattering}, x-ray diffraction data confirm that the series of alloys constructed along the NaRuO$_2$--Na$_2$RuO$_3$ pseudobinary phase diagram are predominantly single phase, with a only a small secondary fraction of Ru metal. In the spirt of phase-boundary mapping \cite{pbmcrawford2018experimental,pbmohno2017achieving,pbmohno2018phase,pbmortiz2019carrier}, this impurity was intentionally introduced to pin the samples to the Ru-rich edge of the single-phase region. Significant changes in peak positions and the corresponding lattice parameters (Fig.~\ref{fig:Vegard}) are clearly observed in the x-ray scattering measurements.

A summary of the changes in the crystallographic parameters accompanying the transition from NaRuO$_{2}$ to Na$_{2}$RuO$_{3}$ is presented in Fig.~\ref{fig:Vegard}. The cell volume increases both monotonically and linearly from NaRuO$_{2}$ ($x$=0) to Na$_{2}$RuO$_{3}$ ($x$=1), consistent with Vegard's Law. This serves as confirmation of a solid solution, and further highlights the propensity for the formation of Na$_\text{Ru}$ antisite defects in NaRuO$_{2}$. Unexpectedly, the off-stoichometry of disordered Na$_{2}$RuO$_{3}$ is similarly complex and has the ability to absorb excess Na up to $x$=4/3. Past this point, samples become multiphase and exhibit a mixture of Na-rich Na$_{3+x}$Ru$_{3-x}$O$_6$ and Na$_{3}$RuO$_{4}$. It is interesting to note that the symmetry of Na$_{3}$RuO$_{4}$ (space group $C2/m$) is a subgroup for $R\bar{3}m$ and is structurally similar to NaRuO$_{2}$ and Na$_{2}$RuO$_{3}$ ($e.g.$ 6-coordinate Na/Ru, approximate planes of metal cations).

The volumetric expansion of the lattice observed in Fig.~\ref{fig:Vegard} with additional Na-loading can be rationalized through simple ionic radii arguments. In a 6-coordinate environment, the Shannon radius of Ru$^{3+}$ is 0.68~\AA\ and Ru$^{4+}$ is 0.62~\AA. While excess Na is expected to convert Ru$^{3+}$ to Ru$^{4+}$, the effect of substituting the much larger Na$^+$ (1.02~\AA) on Ru$^{3+}$ dominates. Thus, a general expansion of the lattice is expected as Na$_\text{Ru}$ defects accumulate. 

The \nro~solid solution poses a synthetic challenge, particularly when the stoichiometry of polycrystalline NaRuO$_2$ needs to be tightly controlled. As illustrated in Fig.~\ref{fig:Vegard}(b), the \nro~solid solution creates several large 2-phase (blue) regions where \nro~ is at equilibrium with NaRu$_{2}$O$_{4}$ under O-rich conditions, Ru metal under O-poor conditions, and Na$_{3}$RuO$_{4}$ under Na-rich conditions. Three unique three-phase (gray) equilibria were identified between Na$_{3+x}$Ru$_{3-x}$O$_6$--NaRu$_2$O$_4$--Ru, Na$_{3+x}$Ru$_{3-x}$O$_6$--Na$_2$RuO$_4$--Na$_3$RuO$_4$, and Na$_{3+x}$Ru$_{3-x}$O$_6$--Na$_3$RuO$_4$--Ru. In our experience, the NaRuO$_2$--Na$_2$RuO$_3$ alloy does not readily support off-stoichiometry in the Ru-rich direction beyond NaRuO$_2$. Employing the principles of phase boundary mapping, we would aim to synthesize NaRuO$_2$ under conditions that place it in equilibrium with NaRu$_2$O$_4$ and Ru metal. A convenient metric would be to minimize the cell volume of NaRuO$_2$.

Attempts to make samples in the O-rich region above nominal Na$_{2}$RuO$_{3}$ indicate the presence of \textit{at least one} unknown Na--Ru--O ternary, complicating the mapping process. Although we would na\"{i}vely suspect samples to contain Na$_{27}$Ru$_{14}$O$_{48}$  \cite{allred2011na27ru14o48}, this phase could not be reproduced using the processing techniques described here. Considering the potential complexity in this region of the diagram, we refrain from postulating on the phase equilibria in this region. This is complicated by the existence of the Na$_{3-x}$Ru$_4$O$_9$ solid-solution, creating large swaths of 2-phase regions. Future work will be required to fully understand the O-rich side of the Na--Ru--O phase diagram. 

Regardless of the additional complexities present in the O-rich regime, the isothermal phase diagram presented here establishes a reliable method for Ru-rich processing of NaRuO$_{2}$, minimizing the substitution of nonmagnetic Na$_\text{Ru}$ defects on the Ru triangular lattice. Compositions located in the three-phase NaRuO$_2$--NaRu$_2$O$_4$--Ru Alkemade triangle will reliably produce NaRuO$_{2}$ at the compositional invariant point where the ternary Alkemade triangle adjoins the vertex of the \nro~single-phase region. Tuning the composition to produce NaRuO$_{2}$ at this vertex with minimal contributions from Ru-metal and NaRu$_2$O$_4$ enables stoichiometry control in a system with a complex phase diagram containing volatile elements.

\begin{figure}
\includegraphics[width=\linewidth]{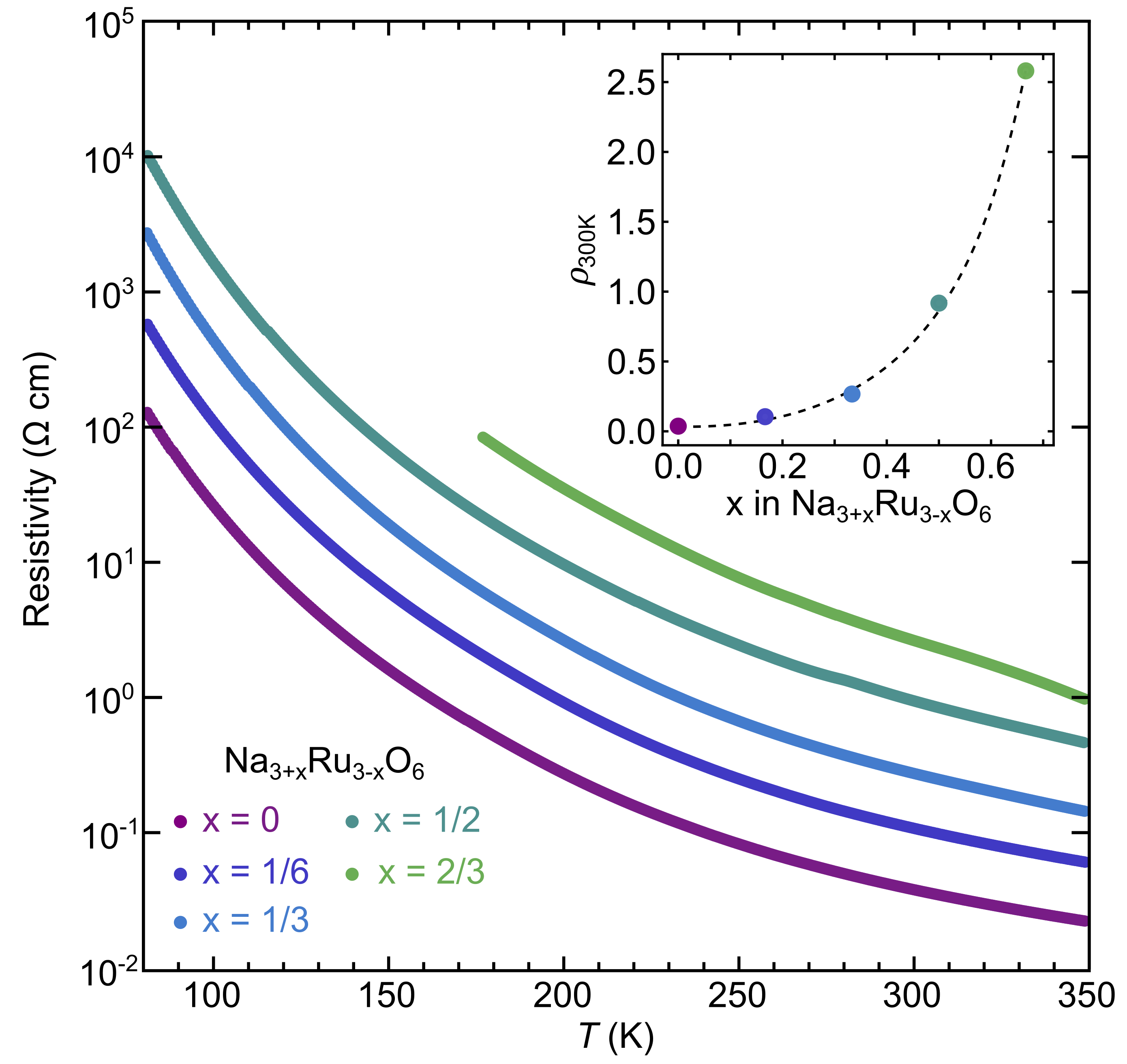}
\caption{Temperature dependence of electronic resistivity of \nro~alloys up to $x=2/3$ is consistent with a lightly doped insulator, with (inset) resistivity increasing exponentially with Na incorporation.}
\label{fig:Transport}
\end{figure}

\begin{figure*}
\includegraphics[width=1\textwidth]{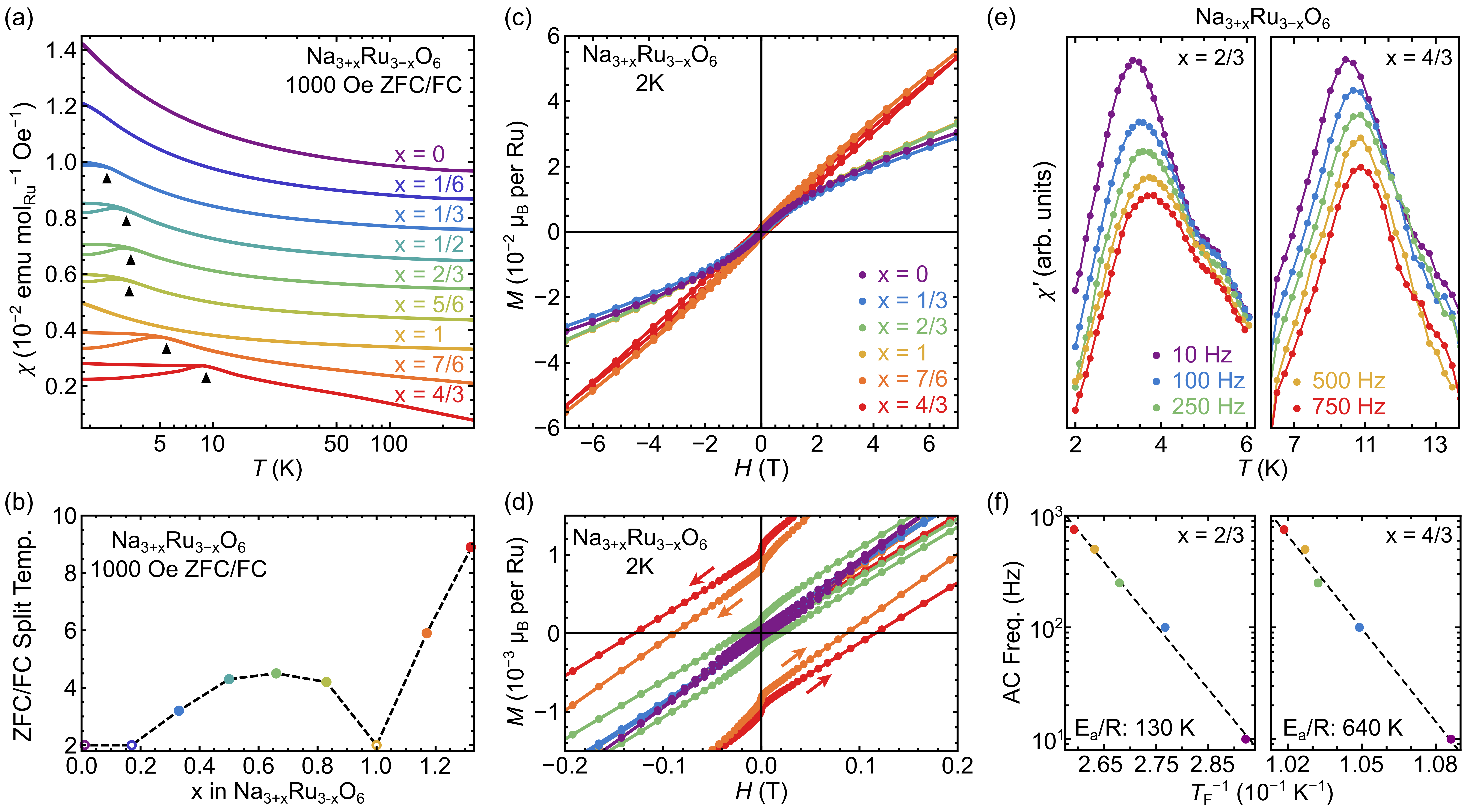}
\caption{(a) Temperature dependence of the ZFC and FC dc magnetic susceptibility for Na$_{3+x}$Ru$_{3-x}$O$_6$ alloys in an external applied field of 1000~Oe. Black triangles denote bifurcation temperatures of the ZFC/FC curves. (b) Compositional dependence of the ZFC/FC bifurcation temperature. Peaking for intermediate compositions, ZFC/FC splitting falls below 2~K for the nominal end members $x$=0 and 1. (c) Field dependence of the dc isothermal magnetization at 2~K with (d) magnified view about $H=0$, highlighting non-zero coercivity for intermediate Na loading. Note that the coercivity vanishes to within the level of background for $x$=0 and 1. (e) Temperature dependence of the in-phase component $\chi'$ of the ac susceptibility in the absence of an external dc field for samples $x = 2/3, 4/3$ with (f) corresponding Arrhenius plot fit to empirical form $f\propto e^{\frac{-E_a}{RT_F}}$.}
\label{fig:Mag}
\end{figure*}

\subsection{Magnetization and Electrical Transport}

Our prior investigation on both the magnetic and electronic properties of stoichiometric NaRuO$_{2}$ identified the system as a magnetic insulator with a quantum disordered ground state \cite{ortizNaRuO2}. Considering that Na$_{2}$RuO$_{3}$ was considered a distinct compound to date, the discovery of the \nro~solid solution should provide an experimental route to exploring the physical properties and possibly unique crossovers ($e.g.$ metal-to-insulator) between the endpoint members. However, literature reports on the magnetic and electronic properties of Na$_{2}$RuO$_{3}$ are varied. Much of the variation stems from the ambiguity whether the ordered or disordered polymorph is present. Even within studies focused predominantly on disordered Na$_{2}$RuO$_{3}$ or mixtures of the ordered/disordered phase, there are conflicting reports. Some works suggest insulating behavior with long-range antiferromagnetic order \cite{Wang14:90,gapontsev2017spectral}, while others report a paramagnetic, moderately correlated electron metal with no observable magnetic excitations \cite{Veiga20:4}.

This lack of consensus on Na$_{2}$RuO$_{3}$ is likely driven by the existence of the \nro~solid-solution. Since Na$_{2}$RuO$_{3}$ is not a line compound, the stoichiometry of a given synthesis is not well-defined. In the case of disordered Na$_{2}$RuO$_{3}$, the majority of samples were produced as a product of decomposition reactions, yielding lattice parameters \textit{a} : [3.11--3.17~\AA] and \textit{c} :  [15.94-16.04~\AA] \cite{mogare2004syntheses,tamaru2013layered,Veiga20:4}. One of the ``hallmark'' features of disordered Na$_2$RuO$_3$ in prior work is the merger of the (101) and (006) peak positions. In good agreement with prior literature, we find that the peak merger occurs with \textit{a}=3.11~\AA~and \textit{c}=15.94~\AA. However, our nominal stoichiometry at that point is only $x$=2/3 instead of $x$=1. This is conceptually consistent with our findings that the Na--Ru--O systems require additional Na and O to compensate for volitility issues. Furthermore, Na incorporation continues well past the point of peak merger -- and well beyond nominal Na$_{2}$RuO$_{3}$ (Fig.~\ref{fig:Vegard}).

The \nro~solid solution presents an opportunity to study the defect-sensitivity of NaRuO$_2$ and the consequence of diluting the Ru-sublattice. We first address the electrical resistivity to determine whether all members of the \nro~solid solution remain insulating, or whether the Na$_\text{Ru}$ defects cause any increase in the free carrier concentration. As illustrated in Fig.~\ref{fig:Transport}, the resistivity at room temperature for many of the series falls within the lightly doped semiconducting regime (10-100\,m$\Omega$-cm), and rises exponentially with decreasing temperature. Both observations suggest that members of the Na$_{3+x}$Ru$_{3-x}$O$_6$ solid solution up to $x$=2/3 are insulators or small-gap semiconductors. 

The isothermal resistivity at 300~K (Fig.~\ref{fig:Transport}(inset)) exhibits an exponential \textit{increase} with Na content, contradicting the most facile defect formation ($e.g.$ $\text{Na}_\text{Ru} + 2\text{h})$ and instead supports the localization of holes via a shift of Ru into a higher oxidation state. The influence of poorly screened, higher charged Ru$^{4+}$ -- coupled with increased alloy/disorder scattering likely contribute to the strong resistivity increases. Potentially more complex compensation reactions such as oxygen vacancies could be present, and more research (e.g. DFT defect studies) will be important for fully understanding the defect energetics in the alloys. We note here that members with higher Na content ($x\geq$1) become progressively deliquescent and will condense atmospheric water on the surfaces, precluding reliable measurement of their resistivity.

The dc susceptibility data for select Na$_{3+x}$Ru$_{3-x}$O$_6$ compositions are plotted in Fig.~\ref{fig:Mag}(a). A manual vertical offset has been introduced to facilitate a visual qualitative comparison, and an unscaled set of magnetization curves is included in the supplementary information for comparison \cite{ESI}. Notably, an onset of irreversibility in the ZFC/FC curves appears in compositions with noninteger $x$.  This irreversibility is absent in the stoichiometric $x$=0 end member above 2 K.  Then, as summarized in Fig.~\ref{fig:Mag}(b), ZFC/FC irreversibility onsets at finite $x$ and increases in temperature as further disorder is introduced.  Near the midpoint between NaRuO$_2$ and Na$_2$RuO$_3$, the irreversibility temperature reaches a local maximum and then begins to decrease again as $x=1$ is approached. In the nominal $x$=1 composition with uniform Ru$^{4+}$ sites, the system naively assumes a $J_\text{eff}=0$ nonmagnetic singlet state and irreversibility vanishes.  With continued Na loading beyond $x=1$, moments are reintroduced and a sharp reemergence of irreversibility occurs. It should be noted that as $x=0,1/6,1$ samples exhibit no discernible splitting by 2~K (though the curvature of $x=1/6$ is suggestive of a splitting proximal to 2\,K), this lower limit on the onset of an irreversibility temperature is denoted as open circles in Fig.~\ref{fig:Mag}(b).

As illustrated in Figs.~\ref{fig:Mag}(c,d), the main qualitative trends presented in Fig.~\ref{fig:Mag}(b) are also reflected in the compositional dependence of the dc magnetization. Compositions with higher irrversibility temperatures exhibit larger coercivity, particularly for those samples where $x>1$ (Fig.~\ref{fig:Mag}(d)).  Irreversibility in FC/ZFC data reflect that local Ru moments freeze, and Fig.~\ref{fig:Mag}(e) illustrates this freezing further in the Na-rich side of the phase diagram with ac-susceptibility measurements over the splitting temperature for $x$=2/3 and $x$=4/3.  The ac-susceptibility data reveal a clear frequency-dependence associated with local moment freezing in both samples.

 High activation energy barriers are obtained for both $x$=2/3 and $x$=4/3 (130\,K and 640\,K, respectively) when analysis is performed solely using the Arrhenius model. Attempting to utilize alternative models (e.g., Vogel-Fulcher) yield similarly unusual characteristic times. Understanding the freezing dynamics in the Na$_{3+x}$Ru$_{3-x}$O$_6$ alloys will require more detailed measurements and neutron scattering measurements on single crystals. However, qualitatively these results demonstrate that the chemical and valence disorder imparted by Na$_\text{Ru}$ defects throughout the magnetic sublattice acts to initiate freezing, consistent with our prior work suggesting that NaRuO$_2$ possesses a quantum disordered ground state \cite{ortizNaRuO2}.

It is worth stressing here that even in the nominal $x=0$ composition, a low-temperature cusp appears in the ac-susceptibility below 2 K \cite{ortizNaRuO2}.  Near 1.7 K, signs of partial moment freezing were observed, indicating a weak spin freezing transition and crossover in the low frequency spin dynamics.  This crossover/partial freezing is likely driven by a small percentage of remnant Na defects ($\approx 1$ $\%$).  This is consistent with the amplification of the freezing onset upon the intentional introduction of additional Na defects along the solid solution line between NaRuO$_2$ and Na$_2$RuO$_3$.

\section{Conclusions} 
Motivated by the need to control and understand defect relationships in the Heisenberg-Kitaev candidate material NaRuO$_{2}$, we studied the chemical potential phase space surrounding NaRuO$_{2}$. We discovered the existence of a full solid-solution \nro~between NaRuO$_{2}$ ($x$=0) and disordered Na$_2$RuO$_3$ ($x$=1). While resistivity measurements demonstrate that all members of \nro~are insulators, increased Na-incorporation into the alloy results in a glass-like freezing of local Ru moments between stoichiometric endpoints. At small $x$, this is conceptually consistent with moment dilution/induced freezing on a highly frustrated Ru$^{3+}$ sublattice. Our study provides key information needed to control chemical disorder and off-stoichiometry in the Heisenberg-Kitaev candidate material NaRuO$_2$.



\section{Acknowledgments}
We acknowledge fruitful conversations with A.~A.~Aczel, G.~Pokharel, and A.~R.~Ericks.  This work was supported by the US Department of Energy (DOE), Office of Basic Energy Sciences, Division of Materials Sciences and Engineering under Grant No. DE-SC0017752.  B.R.O. and P.M.S. both acknowledge financial support from the California NanoSystems Institute through the Elings Fellowship program.  The research made use of the shared facilities of the NSF Materials Research Science and Engineering Center at UC Santa Barbara (DMR- 1720256). The UC Santa Barbara MRSEC is a member of the Materials Research Facilities Network. (www.mrfn.org). This work also used facilities supported via the UC Santa Barbara NSF Quantum Foundry funded via the Q-AMASE-i program under award DMR-1906325.

\bibliography{NaRuO2_v3}

\end{document}